# Price effects and pass-through of a VAT increase on restaurants in Germany: causal evidence for the first months and a mega sports event

Matthias Firgo[*]

Munich University of Applied Sciences &
CARRI – Center for Applied Research for Responsible Innovation

02/09/2024

This paper analyses the price effects and tax pass-through of a VAT increase from 7% to 19% on restaurant services in Germany as of January 1, 2024. The Synthetic Control Method (SCM) is used to identify the causal effects of this reform using prices of goods and services unaffected by the tax change as a counterfactual for restaurant prices. Immediately in January, 31% of the tax increase was passed on to consumer prices. Pass-through increased to 58% in the following six months, which corresponds to a causal consumer price increase of about 6.5%. The presumed increase in demand for gastronomy services due to hosting the UEFA Euro 2024 tournament did not alter the path of price adjustments compared to previous months.

[*] Munich University of Applied Sciences, Department of Tourism, Schachenmeierstr. 35, 80636 Munich, Germany, Email: matthias.firgo@hm.edu, Phone: +49 89 1265-2151; The author gratefully acknowledges helpful comments from Andreas Steinmayr, Firuze Alpaslan and Szymon Kielar.

## 1. Introduction

Like many countries, Germany introduced temporary measures to mitigate economic shocks during the COVID-19 pandemic. In contrast to other measures, the temporary decrease in VAT for gastronomy services (food and drinks in bars, restaurants, etc.) from the standard rate of 19% to 7% introduced in mid-2020 was extended by the national government until 2023. After positive signals for further and calls for unlimited extension, the government announced in November 2023 that VAT in restaurants would return to 19% by January 2024.

The purpose of the present paper is twofold: First, it contributes to the scarce literature on tax pass-through in hospitality, a topic of considerable importance for policymakers and industry stakeholders in the context of post-pandemic recovery. It is based on the Synthetic Control Method (SCM), a prominent tool in economics and political science to quantify the causal effects of a policy intervention. Second, it studies the evolution of pass-through rates in the short and medium run for periods of heterogenous consumer demand. While the first months after the tax reform were characterized by rather low macroeconomic dynamics, hosting the UEFA Euro 2024 tournament was supposed to induce a positive demand shock for gastronomy services in early summer due to an influx of international fans and universal broadcasting in restaurants and bars.

## 2. Related Literature

The impact of a change on an indirect tax such as VAT is significantly influenced by the extent to which it is passed on to consumer prices. Microeconomic theory states that the less price-elastic side bears a larger share of the tax burden. If consumers are more price-sensitive relative to the supply side, the resulting fall in demand limits the extent of price pass-through (Fullerton & Metcalf, 2002). VAT pass-through has been thoroughly investigated in empirical studies. In general, empirical results on pass-through rates are highly dispersed (Loretz & Fritz, 2021). Benedek et al. (2020) examine VAT reforms in Europe, finding an average pass-through of 79% for changes in the standard VAT rate. These results are consistent with Büttner & Madzharova (2021), who identify an average pass-through of around 73% across a range of VAT reforms. However, for a broader set of commodities, EU countries, and years, Benzarti et al. (2020) on average identify low but asymmetric 34% for increases and 6% for decreases of VAT rates.

For restaurant services, evidence on the impact of a change in the VAT rate on prices is limited. Benedek et al. (2020) find a low 25% pass-through in France and no pass-through at all in Belgium. This contrasts with the results of a full pass-through of a VAT reduction for food retail in Norway (Gaarder, 2019). Fuest et al. (2024) examine the temporary VAT reduction in food retail in Germany between July 1, 2020, and January 1, 2021. For food and non-alcoholic beverages, the reduction was passed on at a rate of 80%. For alcoholic beverages and tobacco, the pass-through was only 20% to 30%. The difference is the result of different price elasticities of demand as well as limited possibility of stockpiling due to perishability for many food products.[1]

## 3. Identification Strategy and Data

The study design exploits the fact that the tax change only affects restaurant services but no other goods or services, and that it was announced at rather short notice on November 17, with the change being in affect already on January 1. Thus, the news provided an exogenous asymmetric shock to the industry with little room for anticipation effects in prices. In addition, the reform only affects a small proportion of total private consumption and a significant share of consumption of restaurant services comes from foreign visitors. Therefore, indirect income effects can be regarded as negligible (Loretz & Fritz, 2021).

Monthly consumer price indices that are representative at the national level are available from the German Federal Statistical Office (Destatis) at a highly disaggregated level of

---

[1] In a related analysis for Germany, Montag et al. (2020) find fast but incomplete pass-through of a temporary VAT reduction on fuel, with substantial heterogeneity between fuel types.

Individual Consumption by Purpose (COICOP) classification. Restaurant services affected by the tax change can be isolated from gastronomy services that saw no change in VAT at the COICOP 5-digit level (COICOP 11.11.1 - Food and drinks in restaurants, cafés, bars, etc.). In contrast, price indices in the pool of donor COICOP divisions serving as control group are selected at the 3-digit level. This is to avoid unstable trends of more disaggregated indices that would impede proper identification.

Only COICOP divisions unlikely to be affected by spillovers from changes in consumption patterns following the change in VAT on restaurants are used to construct the counterfactual for restaurant prices. Several types of COICOP divisions are excluded from the donor pool: i) other hospitality services; ii) retail food & beverages; iii) services with a high share of public or non-profit providers such as transport, postal and courier services, education, and social services; iv) services with predominantly annual rather than constant price changes (insurances, public utilities, housing rents). Accordingly, 25 COICOP 3-digit divisions remain in the final donor pool to construct the synthetic time series for restaurant prices through SCM (see Abadie, 2021, for details).[2]

November 2023, the month of the announcement, marks the beginning of the treatment period rather than January 2024 because restaurants may have changed prices immediately. The 12 months before the announcement are defined as the pre-treatment period (11/2022 – 10/2023). The corresponding 12 monthly price index values are used as control variables as suggested by Kaul et al. (2021) in the absence of alternative control variables (that would need to vary at the level of COICOP divisions). The analysis is carried out using the standard *synth2* command for SCM in Stata.

---

[2] Note that using other price indices for the same country rather than the same price index for other countries reduces the risk of miss-specification due to differences in national inflation rates and consumption dynamics. This is particularly relevant for the period studied, during which growth and inflation rates diverged substantially between EU countries.

Figure 1 illustrates the price index of restaurant services (COICOP 11.11.1) and the total consumer price index (CPI) between 11/2022 and 07/2024 (i.e., the most recent period available when the paper was completed). The vertical line highlights the month before the announcement of the upcoming change in VAT rate (10/2023). As Figure 1 illustrates, the hike in CPI that Germany shared with most of Europe in 2022 and 2023, came to an end in late 2023. While prices of restaurant services correlated very strongly with CPI until that period, the change in VAT in the former induced a strong increase in prices that continued during the first half of 2024.

Figure 1 – Restaurant prices and total consumer price index in Germany

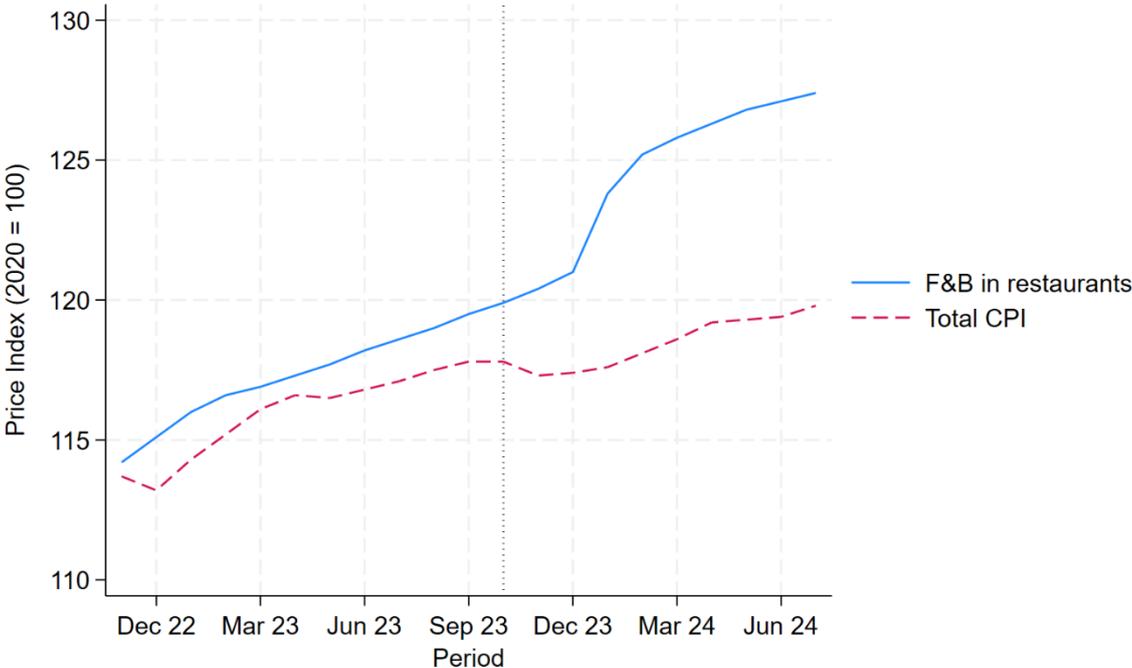

Source: Destatis, own calculations; F&B … food and beverages.

## 4. Results

Figure 2 compares the actual restaurant prices of Figure 1 with a synthetic price index at 7% VAT for these services based on SCM (short-dashed line). The long-dashed line illustrates the corresponding full-pass-through scenario at 19% for the period after the VAT increase. Actual and synthetic prices already deviate in November and December 2023, indicating anticipatory price increases by restaurants between the announcement and the enforcement of the VAT change. The gap increases substantially in January 2024 and continues to grow until the most recent period July 2024. The numbers in Figure 2 indicate the resulting pass-through rate, which was at 31.1% in January and steadily increased to 58.2% in July.

Figure 2 – Actual and synthetic developments of restaurant prices

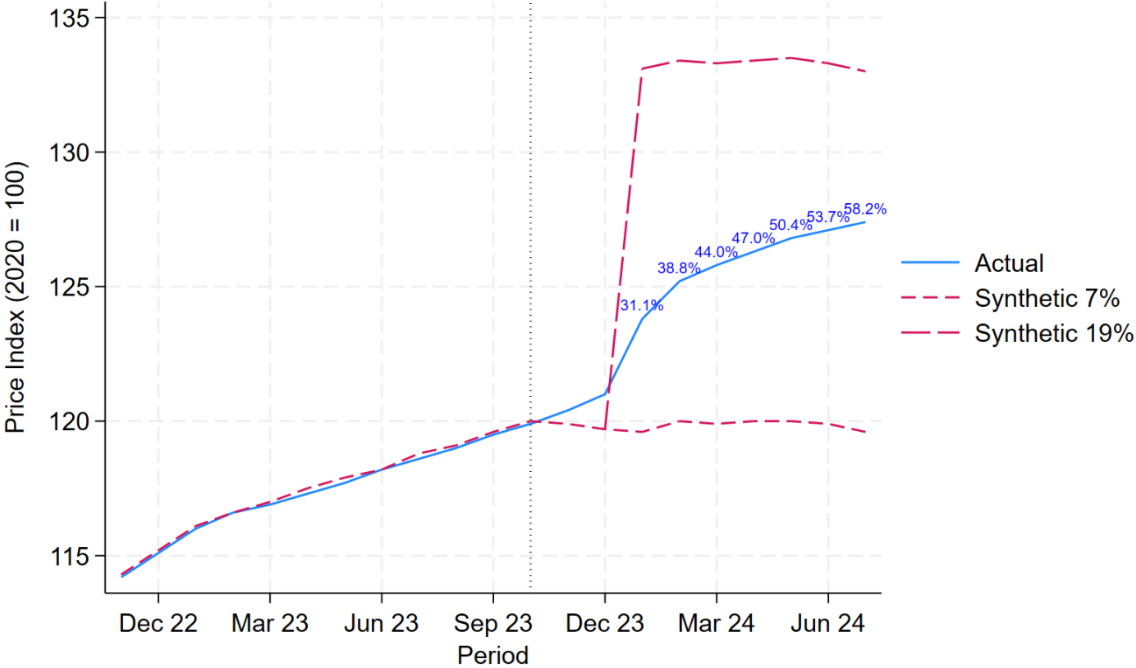

Source: Destatis, own calculations; Percentages reflect the corresponding pass-through rates.

Germany hosted the European Football Championship (UEFA Euro 2024) between June 14 and July 14. Thus, prices for June and July cover a period characterized by an influx of international football fans and numerous large- and small-scale public screening events. While this mega event was expected to imply a positive demand shock for gastronomy services, Figure 2 suggests that the path of tax pass-through did not change notably compared to the first five months of 2024.

Figure 3 – Robustness: Treatment effect and leave-one-out-test (LOO)

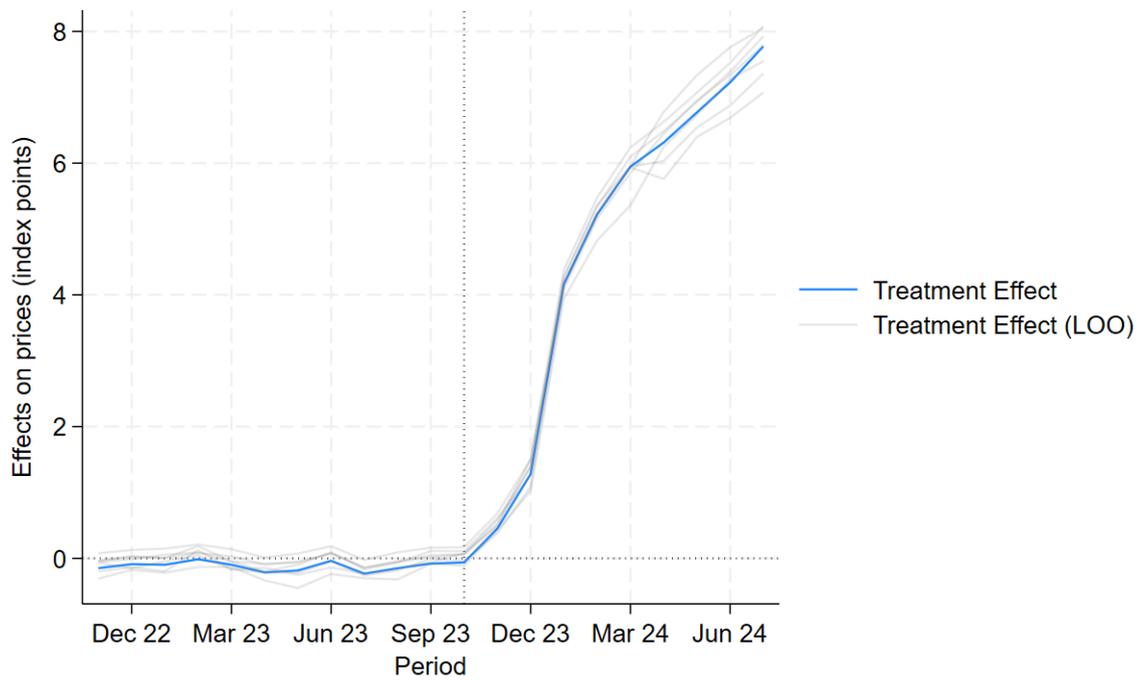

Source: Destatis, own calculations.

Figure 4 – Robustness: Treatment effect and placebo test

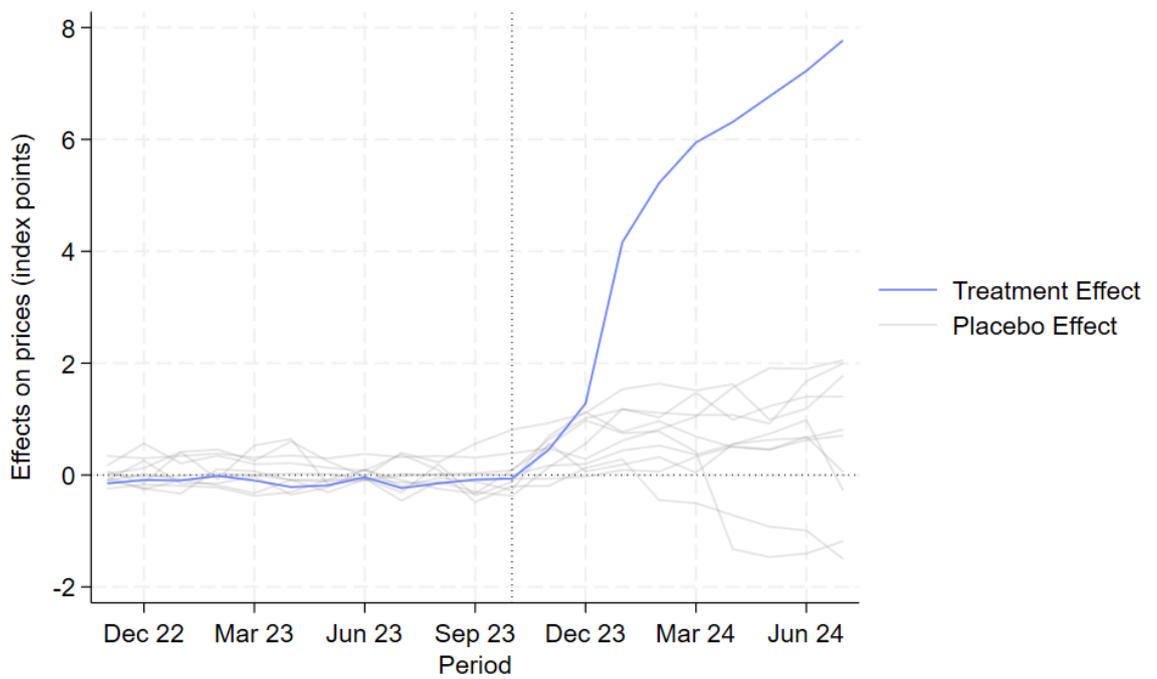

Source: Destatis, own calculations.

Two tests (see Abadie, 2021, for details) are carried out to ensure that the results are not driven by the specification and functional form of the SCM analysis. Leave-one-out estimates (Figure 3) show a high degree of robustness of the actual treatment effect of up to almost 8 index points – a causal price increase of roughly 6.5% by July. In this test the analysis is carried out repeatedly while iteratively leaving out one COICOP division having a positive weight in the synthetic cohort of the main specification. A placebo test (Figure 4) plots the treatment effect against fake treatments for COICOP divisions in the donor pool that were in fact untreated. As revealed, the treatment effect obtained for restaurant services is far outside the range of placebo treatments. This further suggests a causal effect related to the change in VAT, rather than noise in the data constructing the synthetic counterfactual.

## 5. Conclusions

These results indicate that restaurant prices began to rise compared to the synthetic counterfactual immediately following the announcement of the upcoming VAT increase. Still, in January only about 30% of the new tax burden was passed on to consumers. Pass-through increased steadily to almost 60% in July, corresponding to a causal consumer price increase of about 6.5%. However, hosting the UEFA Euro 2024 tournament between mid-June and mid-July did not seem to affect the path of price adjustments. Given the expected low price elasticity of demand for gastronomy services among football fans during such an event, this result suggests substitution effects within gastronomy services and/or a deterrence in consumption among those not interested in football. A more detailed analysis of the impact of the tournament on the various segments of the hospitality sector and on host vs. non-host cities would provide valuable in-depth insights in future research.

While the analysis illustrates monthly dynamics, the identification of long-run equilibrium effects would require a longer observation period. However, the results provide important insights on the distributional effects of the tax change. It shows that the consumer side bears the larger share of the additional tax burden, although firms are unable to pass-through the full burden. Pass-through is higher than in previous studies on France and Belgium, suggesting a relatively inelastic demand for restaurant services in Germany and the post COVID-period.